\documentclass[journal]{IEEEtran} 

\usepackage{color,soul}

\usepackage{xcolor}
\usepackage{graphicx}
\usepackage{lipsum}
\usepackage{caption}
\usepackage{amsthm}
\usepackage{enumitem}
\usepackage{graphicx}
\usepackage{hyperref}
\usepackage{wrapfig}
\usepackage{multirow}
\usepackage{etoolbox}
\usepackage{booktabs}
\usepackage{multirow}
\usepackage{etoolbox}
\usepackage{booktabs}
\usepackage{float}
\usepackage{stfloats}
\usepackage{algorithm}

\usepackage{cite}
\usepackage{amsmath,amssymb,amsfonts}

\usepackage{lipsum}
\usepackage{multicol}

\usepackage{algpseudocode}
\usepackage{graphicx}
\usepackage{textcomp}
\usepackage{subfigure}

\usepackage[para,online,flushleft]{threeparttable}

\usepackage[compact]{titlesec}
\titlespacing{\section}{1pt}{*+1}{*+1}
\titlespacing{\subsection}{1pt}{*+1}{*+1}
\titlespacing{\subsubsection}{1pt}{*+1}{*+1}

\usepackage{enumitem}
\setlist{nolistsep,leftmargin=*}
\setlist{nolistsep}

\begin{document}

\title{
    HMT: A Hardware-Centric Hybrid Bonsai Merkle Tree Algorithm for High-Performance Authentication
}

\author{Rakin Muhammad Shadab, Yu Zou, Sanjay Gandham, Amro Awad \& Mingjie Lin}
\maketitle


\begin{abstract} 

    Merkle tree is a widely used tree structure for authentication of data/metadata in a secure computing system. Even though recent state-of-the art secure systems use MAC based authentication to protect the actual data, they still use smaller-sized MT, namely Bonsai Merkle Tree (BMT) to protect the metadata such as encryption counters. Common BMT algorithms were designed for traditional Von Neumann architectures with a software-centric implementation in mind, hence they use a lot of recursions and are often sequential in nature. However, the modern heterogeneous computing platforms employing Field-Programmable Gate Array (FPGA) devices require concurrency-focused algorithms to fully utilize the versatility and parallel nature of such systems. The predominantly recursive and sequential nature of the traditional BMT algorithms make them  challenging to implement in such hardware-based setups. Our goal for this work is to introduce HMT, a hardware-friendly BMT algorithm that enables the verification and update processes to function independently and provides the benefits of relaxed update while being comparable to eager update in terms of update complexity. The methodology of HMT contributes both novel algorithm revisions and innovative hardware techniques to implementing BMT. We use a partitioned BMT caching scheme that allocates a separate write-back cache for each BMT level, thus allowing for low upper bounds for dirty evictions compared to the traditional BMT caches and providing support for the implementation of existing BMT algorithms, especially lazy update on FPGA-based hardware. Then we introduce the aforementioned hybrid BMT algorithm that is hardware-targeted, parallel and relaxes the update depending on BMT cache hit but makes the update conditions more flexible compared to lazy update to save additional write-backs. Deploying this new algorithm, we have designed a new BMT controller with a dataflow architecture, speculative buffers and parallel write-back engines that allows for multiple concurrent relaxed authentication on-flight which was not possible with the conventional lazy algorithm. Our empirical performance measurements have demonstrated that HMT can achieve up to 7x improvement in bandwidth and 4.5x reduction in latency over baseline in subsystem level tests. In a real secure-memory system on a Xilinx U200 accelerator FPGA, HMT exhibits up to 14\% faster execution in standard benchmarks compared to state-of-the art BMT solution on FPGA.  

\end{abstract}



\maketitle

\section{Introduction}
Due to the massive advancements in fields like cloud computing and blockchain
technology, data privacy and security in local memory and/or servers becomes 
critically important nowadays. 
Recent rise in cyber and hardware-based attacks makes it crucial to
employ state-of-the art attack mitigation and defence mechanisms in both
general-purpose and embedded system scenarios. 
There are several adversarial 
methods that an attacker can use to gain access to information of unsuspecting
users. One of the well-known memory based attacks is a data replay attack where
an intruder can replace some blocks of data and/or metadata with a previously
used redundant block and subsequently utilize this exploit to obtain extra information
on the access pattern~\cite{elbaz2009hardware}. 
As such, mitigating this type of attacks against data privacy and data security
stipulates a strong authentication mechanism.

Generally data or plaintext is encrypted in a secure environment to protect against data confidentiality-based attacks\cite{henson2014memory}. However, even the encrypted data will still be vulnerable to aforementioned memory replay attacks~\cite{elbaz2009hardware}. Therefore, 
strong integrity authentication methods must be used in a secure computing system in order to meet the standard performance and security requirements. Prevalently, Merkle tree or hash trees have been used to authenticate data and/or tagged metadata as they provide robust security against malicious replays~\cite{rogers2007using, henson2014memory}. More recently, 
separate MAC based verification algorithms have started to be adopted for data authentications, which allows for the use of a smaller sized Bonsai Merkle Tree (BMT) to protect only the metadata such as
encryption counters~\cite{rogers2007using}. However, standard BMT update algorithms were originally developed to be used in general purpose computing systems, therefore often containing a large number of recursive function calls~\cite{gassend2003caches}. Unfortunately,
although fairly easy for software-based implementations, these {\em recursive} and {\em sequential} functions are generally not suitable to be implemented With Register Transfer logic (RTL) on FPGA devices deployed in heterogeneous and embedded computing environments. Also, the highly sequential nature of these techniques limits the potential performance improvements provided by such systems. This problem is very relevant since the use of FPGA devices in different domains has been on the rise in recent years due to their versatility, performance benefits, low power requirements and support for heterogeneous computing. With the rapid emergence of FPGA devices for security and accelration in cloud and blockchain applications \cite{bobda2022future, wang2020fpga, oh2021meetgo} and state-of-the art FPGA-driven secure embedded platforms such as ARES~\cite{zou2022ares}, mitigating these two implementation challenges become ever more urgent because of the necessity of using BMT-based authentication in 
many mission-critical heterogeneous and embedded computing systems.

Technically, there are two most common methods of updating a BMT: either eagerly updating all the tree nodes up to the top (with or without a write-through/write-back BMT cache) or just updating the cached nodes in an intermittent manner (with a write-back BMT cache). It is important to note that the counter read/verification operation remains identical across both of the update methods and the authentication process terminates once a cached node is encountered. The use of a BMT cache, while improving performance, makes the hardware implementation of these techniques more challenging due to the additional constraint of recursively verifying uncached tree nodes. Even though the first method, termed eager update, suffers from recursive nature of the verification and update, it can still be implemented on hardware in its native form without a cache by unrolling the recursions. However, the second update technique, known as lazy update ~\cite{zubair2019anubis}, necessitates the use of write-back BMT caches to store the tree nodes. Traditional BMT cache is usually shared between all the BMT levels. In the case of lazy update, such unified cache structure can cause very high number of evictions since a BMT node from any level can evict a node from any other level. This can yield additional complexities since there might be dirty evictions from the cache and each of these evictions can potentially cause further evictions due to extra authentication steps involved with write-backs~\cite{gassend2003caches}. As a result, there might potentially be significant number of supplemental recursive chains. 

Therefore, it is very challenging to implement an unrolled version of lazy BMT update method on hardware using RTL logic since the upper bound for recursions is usually high, not fixed and it will vary with BMT cache size. However, despite its complex nature, lazy update method usually saves a lot of memory access and reduces probable authentication overhead during writes thanks to its relaxed update philosophy and provides potential performance improvements over eager update for write-dependent scenarios. Therefore, it is desirable to use it in the systems where real-time performance is of critical concern. The only effective method to use lazy BMT propagation technique on FPGA with traditional BMT caching is to use an embedded/soft-core processor as a controller to host a software-based high-level implementation of the algorithm. However, due to the overheads of the internal instructions of a processor, such a system will have huge performance degradation compared to a native RTL-based version. While using per-level caching similar to ARES~\cite{zou2022ares} makes an RTL implementation of lazy BMT feasible, its algorithmic bottlenecks still limit the exploitation of concurrency on an FPGA-based design.

We claim the following contributions in this paper: 
\begin{itemize}
\item We talk about the difficulties of unbounded recursions in intermittent BMT updates and introduce {\em a parallel cache structure} that allots a dedicated BMT cache to each level of the tree, consequently limiting the upper bound of the aforementioned recursions and making it feasible to implement relaxed update BMT techniques in RTL. 
\item
Leveraging this new cache
subsystem, We present a hardware-friendly hybrid BMT mechanism called HMT that maintains
the propagation complexity of eager update but relaxes the update conditions
even further than lazy update to solve the issue of additional evictions and
write-backs and allows for only one dirty eviction per tree level in each
authentication. 
\item 
Next, specifically for this hybrid algorithm,
we develop a dataflow
architecture that enables simultaneous processing of multiple in-flight
verification requests which was not possible with traditional BMT algorithm. The architecture, while resembling HERMES~\cite{zou2021hermes} work, handles write requests differently and adds write-back engines to handle dirty evictions for relaxed updates.
\item
The HMT subsystem is directly compared against another intermittent BMT logic (lazy update) in isolated testing scenarios and also pitted against a state-of-the art eager update implementation (HERMES\cite{zou2021hermes}) in a real secure embedded memory-based system with synthetic benchmarks, demonstrating its bandwidth and latency advantages in both experimental setups on Xilinx U200 FPGA platform. The subsystem level tests depict 4.5x latency reduction and 7x uplift in bandwidth over traditional lazy update whereas the synthetic benchmarks in integrated system level tests show up to 14\% better execution performance over HERMES.
\end {itemize}

\section{Technical Background}

The primary idea of Merkle tree-based verification was introduced in a work of Leslie Lamport~\cite{lamport1979constructing, szydlo2004merkle}. Consequently, this idea was extended upon to create a binary tree where the tree leafs are constructed from an irreversible function applied on its children~\cite{merkle1987digital, szydlo2004merkle}. Usually a cryptographic hash function is used for this purpose~\cite{gueron2016memory}. Even though a Merkle tree can effectively mitigate different memory-based integrity attacks including splicing, spoofing and replay, its huge memory overhead of has been a major obstacle for its widespread adoption for the use-cases where data authentication is of paramount importance~\cite{elbaz2009hardware}. This problem was solved by the proposal of a smaller tree called Bonsai Merkle Tree (BMT) that leverages counter-mode encryption technique where the encrypted data would be protected by a hash-based MAC (HMAC) and the tree only needs to protect encryption metadata~\cite{rogers2007using}. Another variant of MT, known as parallel/SGX-style tree or Tree of Counters (ToC) yields higher level nodes using a combination of encryption counters and children nodes and supports parallel update operations\cite{alwadi2020phoenix, zubair2019anubis}. However, ToC implementation typically faces issues with parallel operations in a real-time system due to high overheads and makes the design considerably more complex than BMT in recovery-critical systems\cite{han2021dolos}. Merkle tree and similar integrity trees have been used in different domains for data/metadata authentication or integrity verification. Such tree structures have also been used in for message authentication in blockchain implementations and attack prevention in smart grid systems~\cite{lin2017survey, dattani2019overview, li2013efficient}. There have been notable work with MT-based integrity check in fields like wireless networks~\cite{sun2019non, berbecaru2008performance} and cloud computing~\cite{mao2017position, mohan2020merkle}. The significance of BMT in hardware security is also noteworthy, especially with the emergence of new technologies like Non-Volatile Memory (NVM) where the ability of data retention after a crash allows an attacker to temper the confidentiality and integrity of stored information before the system can fully recover\cite{ye2018osiris, zubair2019anubis, alwadi2020phoenix}. The use of BMT is noticeable in modern FPGA-based reconfigurable systems as a few of the recent FPGA-targeted secure memory platforms have been using eager BMT for integrity authentication of security metadata, while also featuring novel hardware-level optimizations~\cite{zou2019fast, zou2021hermes, zou2022ares}.

\begin{figure*}[h!]
 	\graphicspath{ {./Figures/} }
 	\centering
    \includegraphics[scale =0.5, trim = 0cm 0cm 0cm 0cm]{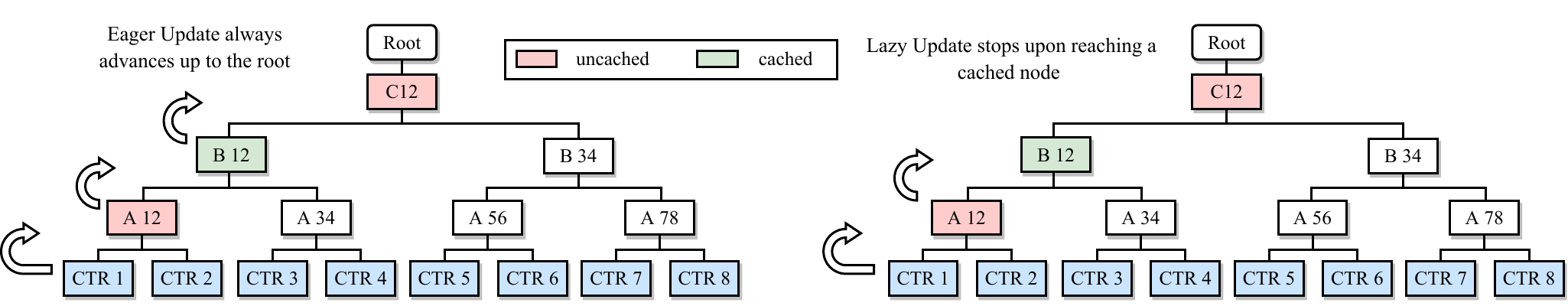}
 	\caption{Eager \& lazy update propagation}
 	\label{eager1}
 \end{figure*}

Utilization of a MT metadata cache can help improving the throughput of a MT subsystem \cite{gassend2003caches, suh2003efficient}. There have been two general methods of updating a BMT during integrity verification, called eager and lazy update process respectively~\cite{awad2019triad}. The eager update scheme propagates the update to every tree node in an update chain all the way up to the root node during each counter modification, therefore requiring a lot of memory access since each parent node needs to be brought to the cache during every operation and each of them needs to be overwritten in the event of a counter write. The lazy update method, on the contrary, only modifies the tree chain up to a cached parent/grandparent node of the data/metadata. The update is not propagated to the higher level nodes in the update chain until the node is evicted from the BMT cache. In other words, reaching a cached node will stop the update process and the node will be used as a temporary root for future updates and authentication of all the children nodes covered by it as long as it resides in the cache. As a result, most of the write requests incurred by sequential real-world applications might generate BMT update chains that only need to advance up to couple of levels above the counter until a cached node is found in the BMT cache. This scenario enables a reduction in the total number of memory access during write operation and also eliminates some of the extra verification requirements caused by additional writes to the uncached upper level nodes. 

BMT integrity authentication \& update requires hash operations on lower level children nodes and form the respective parent nodes in the upper level from the newly generated hash. The encryption counters are hashed and a first level BMT node is engendered from every counter respectively. Every first level node is then hashed and each resultant hash value is then truncated to form a part of the respective parent nodes in the tree. The process is repeated until the tree structure converges to only one parent. This final parent node is truncated to generate the root of tree. The root is always written to a secure on-chip storage as opposed to the regular BMT nodes that are kept in the memory and brought to the BMT cache during ongoing operations.
Specifically,
consider an example (Fig.~\ref{eager1}) where some
encryption counter needs to be verified by a BMT. For every write to a counter,
the eager update process must modify all the parent nodes in the update chain accordingly
including the root, even if one or more nodes in the chain might already be present in the BMT cache. On the contrary, the lazy update scheme (Fig. \ref{eager1}) does not involve
accessing the parent nodes as regularly as eager update and stops as soon as a cached node is found. This scheme requires the use of a write-back metadata cache. 

\section{Threat Model}

The threat model for this work is as follows:
\begin{itemize}

    \item The memory lies in an unprotected region and is susceptible to
        different hardware-based attacks including splicing, spoofing and
        replay attacks. It is also possible to consider other potential attacks
        that can be mitigated by a traditional BMT structure.

    \item Every other component of the system, including the BMT subsystem, possible encryption subsystem and the user logic resides within secure region.

    \item The system emphasizes on real-time
        performance and does not consider crash-consistency or data/metadata recovery.

\end{itemize}

\section{Constraints of Relaxing Updates \& Our Solution} 

The lazy update propagation has the potential to provide better throughput over the eager
update in terms of write performance. However, its possible run-time randomness
makes it unappealing for a real-time embedded system. Due to the mandatory inclusion of a write-back cache, in the worst case, every authentication request might result in several evictions from the cache. In case of dirty eviction, each of these evicted blocks will yield a write-back respectively. However, the nodes need to propagate their updates to the parent/grandparents before the write-back process is complete. Since the update algorithm is recursive in nature, the potential extra verification during these updates might cause even more dirty evictions and as a consequence more write-backs and updates. The worst-case scenario might see all these recursive chains to go on until all the dirty nodes are evicted from the cache, therefore potentially inducing unbounded increase in latency, which is detrimental to real-time computing
performance.
 
\begin{figure*}[htbp]
 	\graphicspath{ {./Figures/} }
 	\centering
 	\includegraphics[scale =0.25, trim = 0cm 0cm 0cm 0cm]{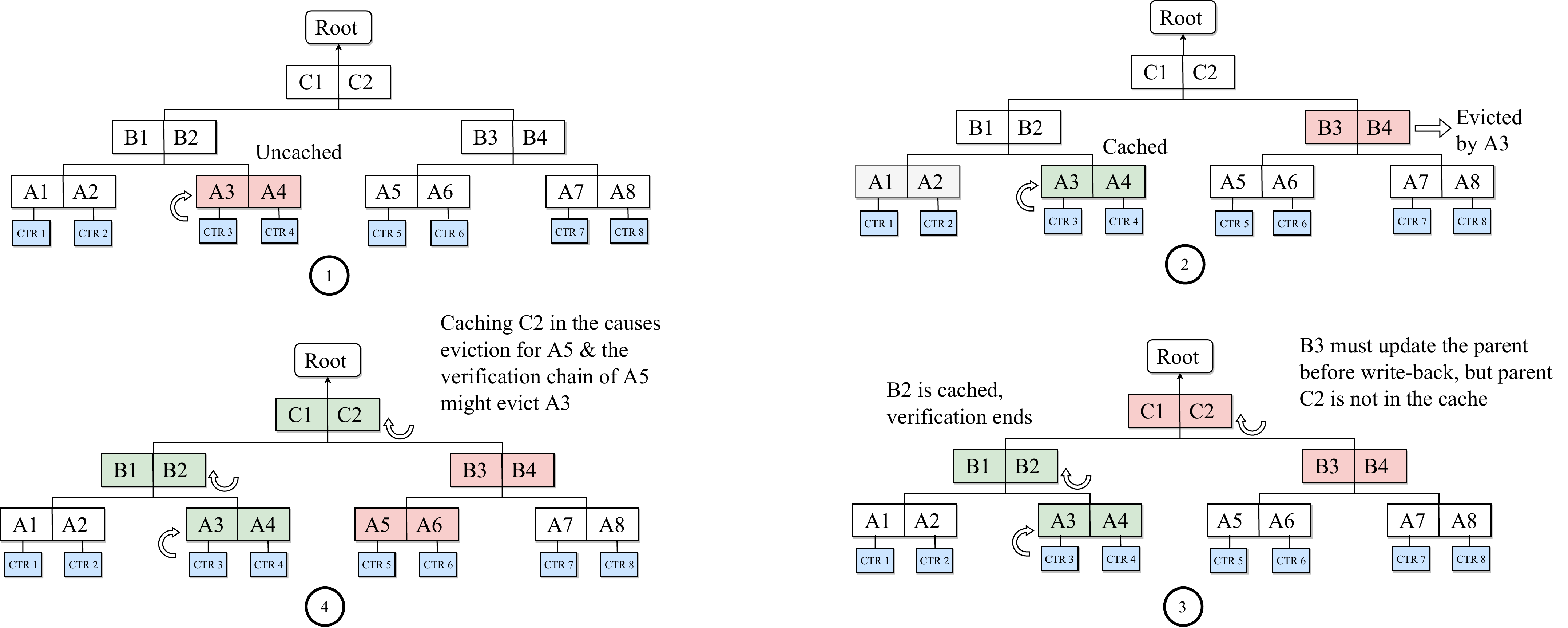}
 	\caption{A depiction of how unbounded evictions might happen during lazy update.}
 	\label{lazy1}
 \end{figure*}

Let's assume that the $\mathtt{CTR3}$ needs to be modified (Fig.~\ref{lazy1}).
The BMT controller will need to fetch its parent node
and update it to propagate this change. After the fetch and subsequent verification of the parent A1, node B3 \& B4 might get evicted. Assuming B3 to be a dirty node, its parent C2 needs to be updated B3 is written back to the memory. Since C2 is not cached, it will have to read together with C1 from memory and verified first. After the verification prcoess, C2 is now put in the cache and as a consequence, A5 is evicted. A5 will need to update its parent B3 before write-back and since B3 does not reside the cache any more, it needs to be read from memory and hence will need to be verified. However, caching B3 again will result in an eviction for A3.
As a consequence, the evictions can continue in an unbounded manner until there is no more dirty eviction. Since theoretically a node can evict any other node from the BMT, the process can go on until all the dirty blocks are thrashed from the cache in the worst-case situation. Therefore, the total number of recursive update/authentication chain due to dirty evictions can be very large in the worst case, equalling to the size of the cache. The worst-case upper bound for dirty evictions will also scale proportionately with the size of BMT cache (upper bound = $S/64$ where $S$ is the size of BMT cache). An increase in the cache size will make the recursive upper bound to go up accordingly and it can be huge for bigger caches. For example, if a 5-level BMT (excluding the counter and root level) with  $S$ = 16KBs of unified BMT cache where each cache block is 64B in size, the worst-case scenrio will see 16KB/64B = 256 worst-case evictions and as many write-backs and recursive chains during one single read/write request. Even though the chances of the worst-case scenario to take place in a real-time system are low, it is still theoretically possible to have notably reduced performance due to the worst-case evictions and any potential lazy BMT authentication setup will need to be designed with this caveat in mind.

 As the evictions are not deterministic by nature, {\em it is highly challenging to unroll the lazy verification algorithm into a hardware-friendly design targeting ASICs and FPGAs}. Perhaps the only realistic solution to use this scheme in
FPGAs in its traditional form is to incorporate the method in a high-level language and run it on a
real-time processor while using the FPGA fabric to implement the supporting elements including BMT cache and hash modules. However, the massive overhead due to
software-based implementation and the internal instructions of the driver processor might result in humongous performance degradation. 

   \begin{figure}[h!]
 	\graphicspath{ {./Figures/} }
 	\centering
 	\includegraphics[ scale= 0.35,trim = 0cm 0cm 0cm 0cm]{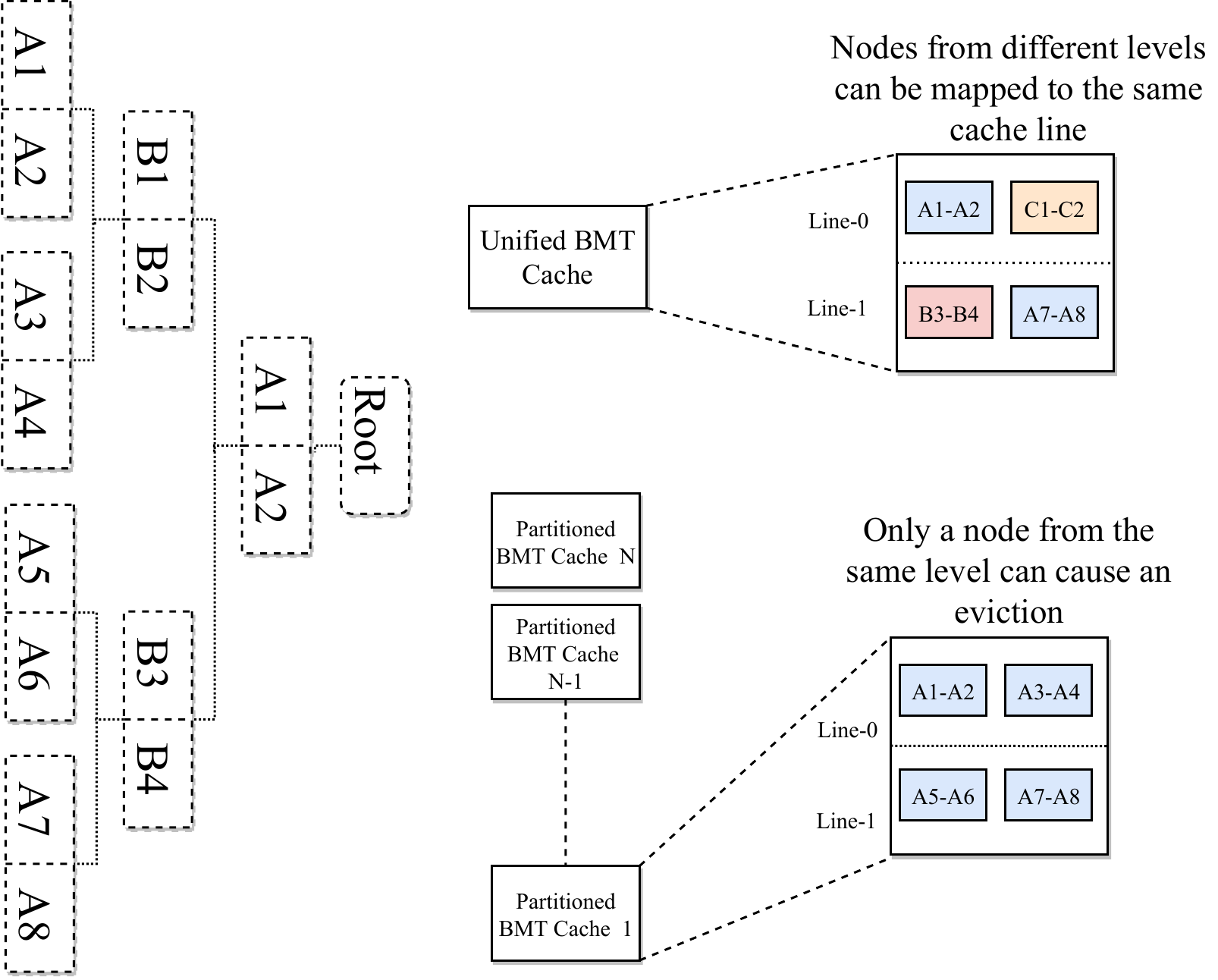}
 	\caption{The use of partitioned cache to reduce worst-case upper bound compared to a regular BMT cache}
 	\label{lazy2}
 \end{figure}

We introduce a new cache architecture to solve the aforementioned challenges
(Fig.\ref{lazy2}). Our caching structure uses multiple parallel caches, each
covering one tree level instead of having one unified cache for all the tree
levels. While ARES\cite{zou2022ares} used similar caching technique, that was intended to only solve the recursive verification during regular integrity check/updates since ARES didn't have to deal with write-back recursion in the first place due to the eager update nature of its BMT. On the other hand, the per-level caching scheme in this work changes the eviction process drastically for the lazy update BMT. Any uncached tree node corresponding to a particular level can now only
evict another node from the same level. Also, an evicted block can no
longer contribute to the eviction of another same-level block during its write-back chain and can only evict a block from the parent level. Furthermore, the solitary child node of the root is assigned a dedicated cache and as a consequence, it will never be evicted. All of the aforementioned features will limit the chances of dirty evictions. As a result, this {\em parallel \& split cache design} imposes a deterministic restriction on the
number of evictions since the critical bound for total number of evictions
during a single counter operation will now only depend on the number of levels
(which is usually fixed for a given data size) and not the size of the cache.
Each cache in this design access the memory independently. Since they are
designated for different levels of the tree, they don't share the same nodes
and as such, there is no conflict among them. 
 
For this proposed cache structure, we calculate the total number of worst-case
write-back recursive chains or in other words, total number of worst-case
evictions due to a counter read or write = $(2^{(N-1)} - 1)$ = $(2^{n} - 1)$
where $n = N-1$ and $N$ is the number of tree level. If we reassess the last
example for a 5-level tree with the new cache architecture, The value
$n$ = $5-1$ = $4$. Therefore, the worst-case upper bound for dirty evictions = ${2^{4}} -
1$ = $15$, demonstrating more than $17x$ reduction compared to the same
example with a unified cache. Unlike the previous case, this upper bound number will remain unchanged for a particular size of BMT even if the cache size changes. This situation allows us to finally unroll all the recursive loops of the chash algorithm in RTL since now we have a
fixed and considerably smaller hard bound for the total number of possible recursive chains during a counter read/write request.

\section{The HMT algorithm: Description and Theoretical Upper Bound}

\begin{algorithm}[h!]
\caption{HMT Algorithm}\label{alg:cap}
\begin{algorithmic}[1]
\footnotesize
\State $L:$ {Current BMT level}
\State $N:$ {BMT height}
\State $M:$ {Verification mask for all levels}
\State $H:$ {Cache hit flag}
\While{Counter channel is not empty}
\If {Dirty eviction from level L}
    \State $D_{L+1} \gets $ {Read data from parent L+1}
    \State $O_{L+1} \gets $ {Read offset from parent L+1}
    \If {$H_{L+1}$ hit}
        \State {Update parent in the cache}
        \State {Write $C_{L}$ to memory}
        \State \Return {Success}
    \Else
        \State {Update parent(s) in the memory until a hit}
        \State {Update cached parent}
        \State \Return {Success}
    \EndIf
\ElsIf {Counter update request}
    \State $D_{L+1} \gets $ {Read data from channel L}
    \State $O_{L+1} \gets $ {Read offset from channel L}
    \If {$H_{L+1}$ hit}
        \State {Update parent in the cache}
        \State \Return {Success}
    \Else
        \State {Update parent(s) in the memory until a hit}
        \State {Update cached parent}
        \State \Return {Success}
    \EndIf
\ElsIf {Counter verification request}
    \State $D_{L+1} \gets $ {Read counter}
    \State $M_{L+1} \gets $ {1}
    \For {$l \gets L$ to $N$}
        \If {$H_{l} \neq 1$}
            \State $D_{L+1} \gets $ {Read data from channel l}
            \State $O_{L+1} \gets $ {Read offset from channel l}
            \State $M_{l} \gets $ {1}
        \Else
            \State $M_{l} \gets $ {0}
        \EndIf
    \EndFor
    \State \Return {VERIFICATION (D,O,M)}
\EndIf
\EndWhile
\State \Function VERIFICATION(D,O,M)
    \For{$l \gets L+1$ to $N-1$ parallel}
        \If{$M_{l}$ and $H(D_{L}) \neq D_{l-1}(O_{l-1})$}
        \State \Return {Error}
        \EndIf
    \EndFor
    \State \Return {Success}
\EndFunction
\end{algorithmic}
\end{algorithm}

  \begin{figure}[h!]
 	\graphicspath{ {./Figures/} }
 	\centering
 	\includegraphics[ scale= 0.75,trim = 2cm 0cm 0cm 0cm]{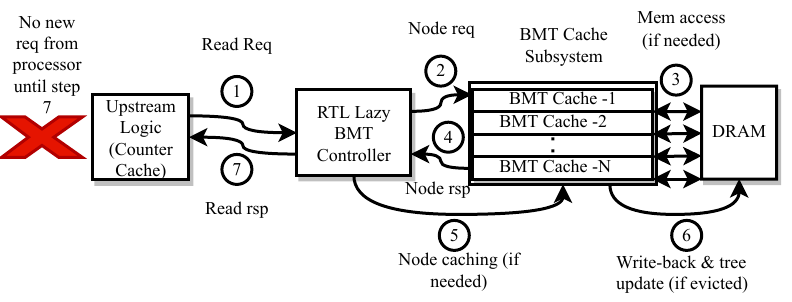}
 	\caption{The sequential workflow of a regular Lazy BMT Controller}
 	\label{lazyl}
 \end{figure}

Even though our proposed cache architecture allows for a theoretical upper
bound for worst-case recursive chains due to dirty evictions that is
independent of the cache size, it shows a non-linear increase in the number of
worst-case write-backs with an increase in number of tree levels. This aspect
of the design makes it very difficult for the MT subsystem to scale up with the
number of levels without making significant changes in the existing design.
Also, even though the base lazy update algorithm can now be unrolled and
realized on hardware with Hardware Description Language (HDL) thanks to our
proposed cache structure, its sequential workflow only enables one integrity
check process at a time. As a result, some blocks in the execution path can
remain idle for most part of the operation even when they can process new
requests. For example, as shown in fig.\ref{lazyl}, a counter cache that holds the encryption counters sends a counter verification/read request when there is a miss in the cache. The counter cache and the rest of the user logic is stalled until the completion of this single counter read/verification request by a sequential lazy BMT controller and no new verification/update requests can be made by the processor or upstream logic until the verification response for the current operation is received. This can create bottlenecks for the system as the rest of the subsystem and even some of the modules within BMT subsystem might remain idle for long periods. To solve these problems, we propose a new hybrid and
hardware-efficient algorithm called HMT. HMT is primarily based on lazy update
technique but differs in two areas - (1) An update to an unverified node no
longer requires a verification of the said node, instead it is updated directly
in the memory. Therefore update and verification functionalities are now
independent from each other, (2) Instead of bringing in an unverified node to
the cache, then verifying and updating it, this new algorithm updates all the
upper level nodes in the memory until it hits a cached node. This process does
not involve bringing any nodes to the cache, therefore it eliminates any
secondary evictions that might have happened due to the write-back chains of
primary evictions. 
 \begin{figure*}[htbp] \graphicspath{ {./Figures/} } \centering
    \includegraphics[ scale = 0.22,trim = 0cm 0cm 0cm
     0cm]{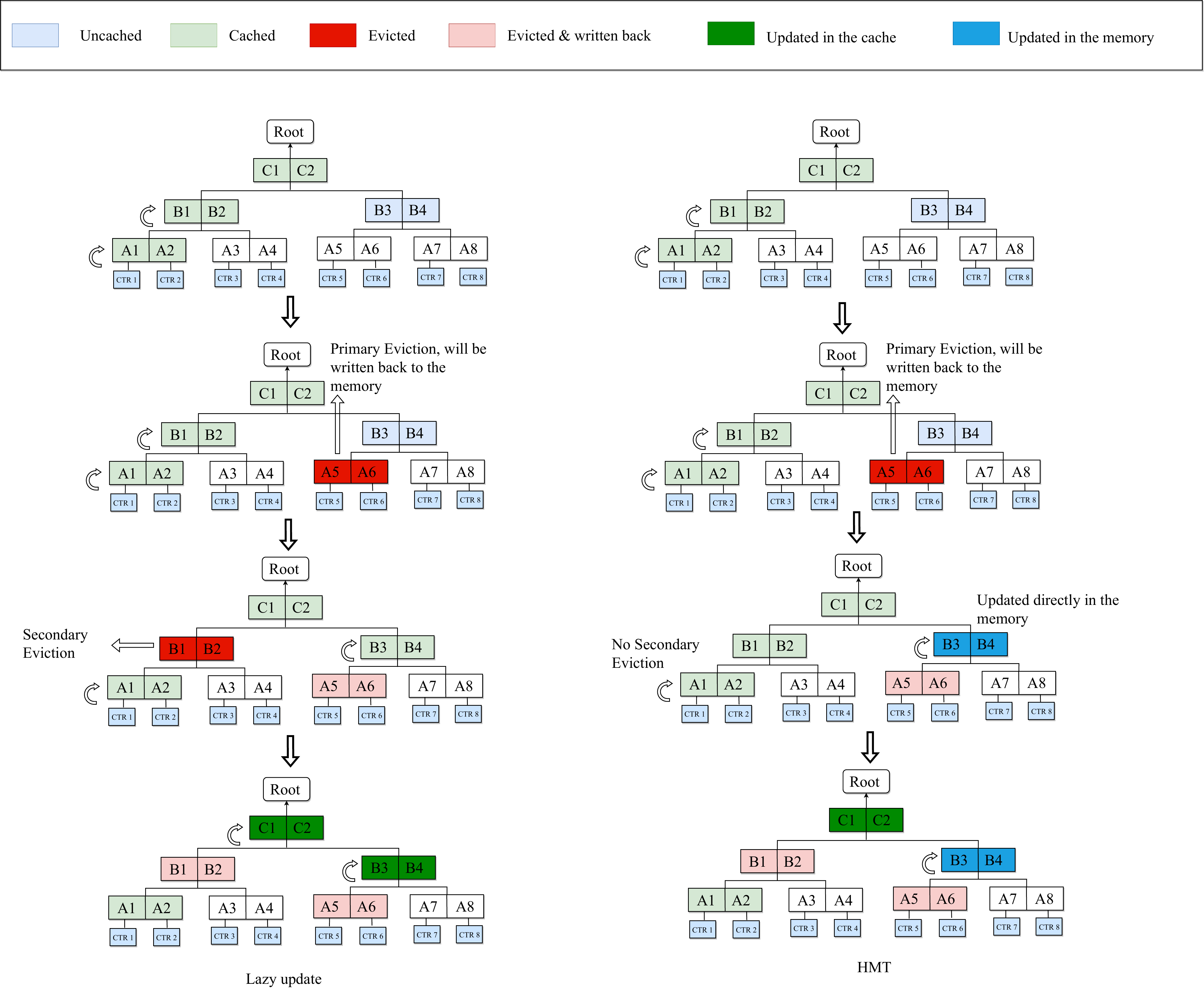}
 \caption{HMT algorithm compared against lazy update algorithm} \label{lazy9}
 \end{figure*} 
 \begin{figure*}[htbp] \graphicspath{ {./Figures/} } \centering
    \includegraphics[ scale = 0.5,trim = 2cm 0cm 0cm
     -1cm]{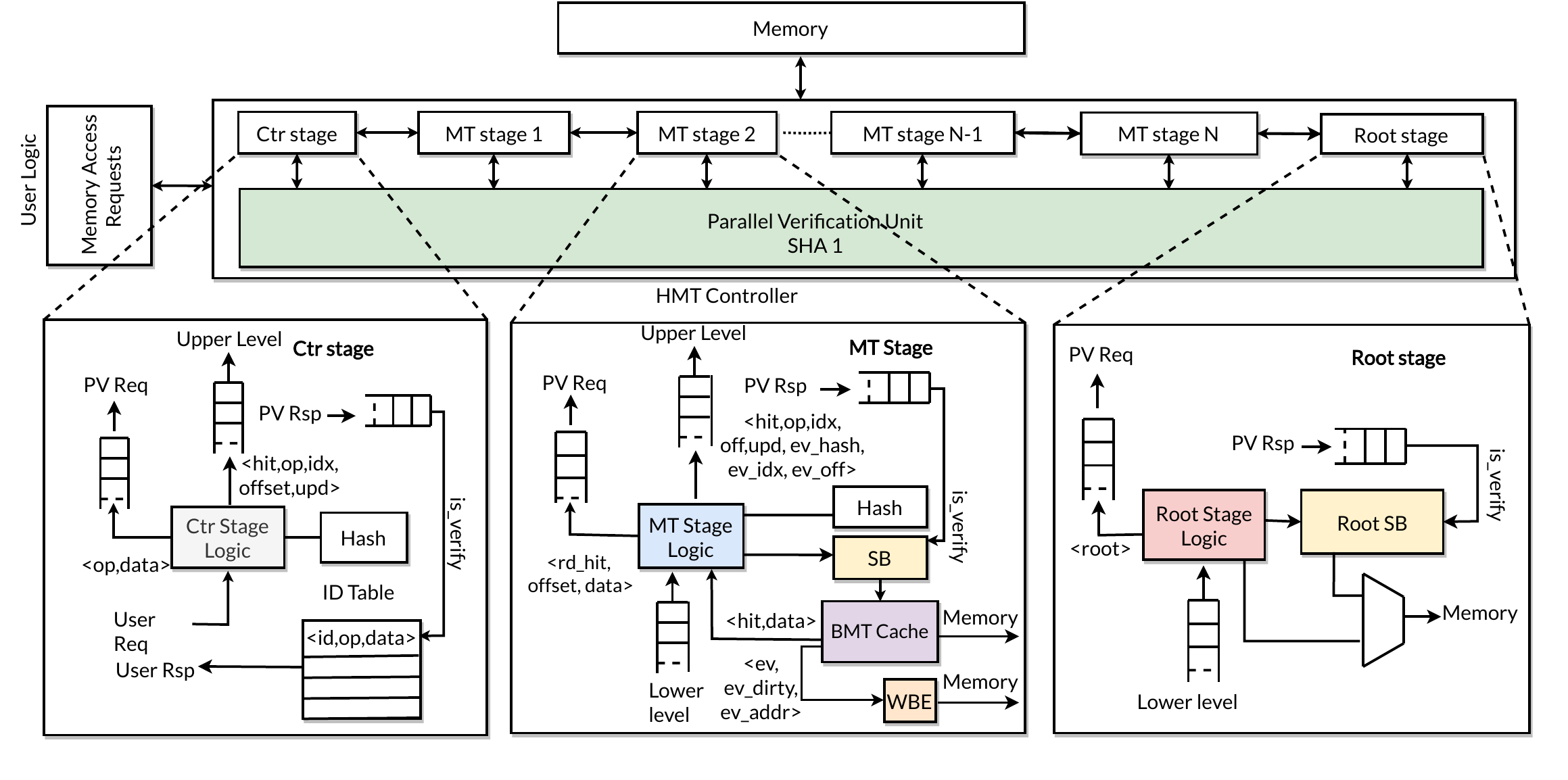}
 \caption{HMT Controller with different compute stages} \label{lazy10}
 \end{figure*}
 Let's consider a verification of counter 1 with our aforementioned partitioned
 cache structure(Fig. \ref{lazy9}). As the request for parent node A1\&A2
 incurs a cache miss, it needs to be verified by comparing its hash against
 cached B1\&B2 node and then stored in the cache . However, bringing A1\&A2 to
 the cache causes a dirty eviction of A5\&A6. Now, according to lazy update
 method, this eviction requires updating the parent node B3\&B4 before the
 evicted node can be written back to the memory. Since B3\&B4 is not cached, it
 is needed to be read from memory, verified and then put in the cache before
 this update. But doing so results in a dirty eviction of B1\&B2. This eviction
 will then have its own write-back chain that involves updating C1\&C2 node. On
 the other hand, our new hybrid algorithm just requires updating the parent
 node, no authentication is needed even if the node doesn't reside in the
 cache. In this case, B3\&B4 is directly updated in the memory. Since the node
 is not brought to the cache, its parent cached node will also be updated. But
 unlike lazy update, there will no longer be any more evictions engendered by a
 previous write-back. Therefore, we will only have the primary evictions caused
 by the original verification chain of a counter. Since each cache level allows
 for only one eviction per operation, the number of worst-case evictions will
 be equal to the number of tree level. Also, updating a counter will no longer
 involve any evictions since it will follow our hybrid update policy.
 Therefore, instead of the overhead of $(2^{(N-1)} - 1)$ like in the case of
 lazy BMT, a N-level HMT-based BMT with our parallel cache subsystem will have
 an upper bound of only N write-backs for counter verification and 0
 write-backs for counter update. This upper bound now follows a very linear and
 direct correlation with the number of BMT levels.

 The write-back mechanism will follow the relaxed update process before writing
 an evicted block into the memory and just like the update, it won't require
 any verification of the parent blocks.  We have also made concurrent
 verification of multiple metadata blocks a part of our proposed algorithm.
 Hence we have a combined framework for verification and a relaxed update
 propagation, both of which are independent processes under our new algorithm.

\section {HMT Controller Design}

The presence of {\em independent parallel authentication} and 
{\em relaxation of the update requirements} 
in HMT algorithm enable us to use a dataflow architecture
for HMT controller. This architecture allows for a separate compute stage for
each of the tree level and every stage can accept a new request as soon as it
finishes the current task. None of the stages will have to wait for the
complete verification of a counter before they can start working on the
integrity check for the next counters. Hence it will enable multiple parallel
counter verification which was not supported by original lazy update
techniques.  Our proposed HMT controller supports three different compute
stages, just like HERMES controller~\cite{zou2021hermes} - one for counter stage, one for root stage and $N$ identical BMT stages
where $N$ is the number of tree levels.
\subsection{Counter Stage}
This Stage uses an FIFO ID table to store counters that are being authenticated/updated. Each entry in the ID table includes additional metadata such as a unique id and the type of operation (read/write). Once an entry in the ID table is verified, it is taken out and the response is sent to the user logic. Unlike HERMES\cite{zou2021hermes} where there is an entry in the table regardless of whether it is a counter verification or update, our design only requires the counter verification/read requests to be registered in the ID table (algorithm \ref{alg:cap2}). Since the counter update process doesn't involve any verification, those operations aren't included in the table and instead the write response is sent to the upstream logic from BMT controller as soon as the update is propagated to a cached node.
\subsection{BMT Stage}
Each of the BMT stages are designed to
perform the same set of tasks - verification, update and write-back. Since
counter and root levels are fixed, a BMT with random size can be easily
supported by simply varying the number of BMT stages. A parallel verification
unit is connected to all the stages (Fig.~\ref{lazy10}). Each of the MT stages
use a speculative buffer (SB) to store the unverified data during an ongoing
verification and the stored node can be used for subsequent operations in the
meantime. These nodes will be written to the cache once the said verification
is done. However, similar to the update mechanism in the counter stage, a write/update request doesn't involve buffering any nodes in the SB. Instead, the write process directly updates the relevant entry in the SB if it resides there. This modified node won't be a new SB entry.

 \begin{algorithm}[h!]
\caption{HMT Controller Workflow}\label{alg:cap2}
\begin{algorithmic}[1]
\tiny
\Function COUNTER STAGE
\While {1}
\If {There is counter request}
    \State {Calculate index and offset of the parent hash} 
    \If {Counter\_read}
        \State{Read ${ctr}$ from memory}
        \State{Make an entry in ${id\_{table}}$ with ${req\_{id}, ctr, off, op =0}$ }
        \State {Forward ${op = 0, ctr}$ to PA}
        \State {Forward ${op = 0, hit = 0, idx, off, upd = 0}$ to MT Stage}
    \Else
        \State{Write ${ctr}$ to memory}
        \State {Calculate hash ${h}$ of ${ctr}$}
        \State {Forward ${op = 0, hit = 0, idx, off, upd = h}$ to MT Stage}
    \EndIf
\EndIf
\EndWhile
\EndFunction
\Function MT STAGE
\While {1}
\If {There is MT request from previous level \& no node write-back request}
    \State {Read entry${op, hit, idx, off, upd}$ from previous level}
    \State {Calculate index and offset of the parent hash}
    \If {$op = 0$} 
        \If {$hit = 1$}
            \State {Forward ${op = 0, hit = 1, idx = 0, off = 0, upd = 0}$ to next level}
            \State {Done}
        \EndIf
        \State {Read $bmt\_node$ from SB or BMT cache}
        \State {$hit\_status \gets $ {$SB\_hit$ or $cache\_hit$}} 
        \State {Forward ${op = 0, hit = hit\_status, idx, off, upd = 0}$ to next level}
        \State {Forward ${hit = 0, off, data = bmt\_node}$ to PA}
        \If {hit = 0}
            \State {Make an entry of $op = 0, idx, dirty = 0, bmt\_node$ in SB}
        \EndIf
    \ElsIf {$op = 1$}
            \State {Read $bmt\_node$ from SB or BMT cache}
            \If {$SB\_hit$ = 1}
                \State {$new\_bmt\_node = upd(bmt\_node, update)$} 
                \State {Over-write the previous entry in SB with $op = 0, idx, dirty = 1, new\_bmt\_node$}
                \State {Forward ${op = 1, hit = 1, idx = 0, off = 0, upd = 0}$ to next level} 
            \ElsIf {$Cache\_hit$ = 1}
                \State {$new\_bmt\_node = upd(bmt\_node, update)$} 
                \State {Update $bmt\_node$ in BMT cache to $new\_bmt\_node$}
                \State {Forward ${op = 1, hit = 1, idx = 0, off = 0, upd = 0}$ to next level} 
            \Else
                \State {$new\_bmt\_node = upd(bmt\_node, update)$} 
                \State {Calculate hash ${h}$ of ${new\_bmt\_node}$}
                \State {Update $bmt\_node$ in volatile memory to $new\_bmt\_node$}
                \State {Forward ${op = 1, hit = 1, idx = 0, off = 0, upd = h}$ to next level} 
        \EndIf
    \EndIf
\ElsIf {There is node write-back request of relaxed update for current level}
       \State {Calculate index and offset of the parent hash}
       \State {Calculate hash ${h}$ of ${ev\_node}$}
       \State {Write $ev\_node$ to $ev\_add$ in the volatile memory}
       \State {Forward $ev\_valid = 1, ev\_hash = h, ev\_idx, ev\_off$ to the next level}
       \State {Go back to MT request}
\ElsIf {There is node write-back request of relaxed update from previous level}
       \State {Calculate index and offset of the parent hash}
       \State {Read $parent\_node$ from SB or BMT cache}
       \If {$SB\_hit$ = 1}
            \State {$new\_parent\_node = upd(parent\_node, update)$}
            \State {Over-write the previous entry in SB with $op = 0, idx, dirty = 1, new\_parent\_node$}
            \State {Forward $ev\_valid = 0, ev\_hash = 0, ev\_idx = 0, ev\_off =0$ to the next level}
        \ElsIf {$Cache\_hit$ = 1}
            \State {$new\_parent\_node = upd(parent\_node, update)$}
            \State {Update $parent\_node$ in BMT cache to $new\_parent\_node$}
            \State {Forward $ev\_valid = 0, ev\_hash = 0, ev\_idx = 0, ev\_off =0$ to the next level}
        \Else
            \State {$new\_parent\_node = upd(parent\_node, update)$}
            \State {Calculate hash ${h}$ of ${new\_parent\_node}$}
            \State {Update $parent\_node$ in volatile memory to $new\_parent\_node$}
            \State {Forward $ev\_valid = 1, ev\_hash = h, ev\_idx, ev\_off$ to the next level}
        \EndIf
        \State {Go back to MT request}
\EndIf
\EndWhile
\EndFunction
\Function ROOT STAGE
\While {1}
\If {There is MT request from previous level}
    \State {Read entry${op, hit, idx, off, upd}$ from previous level}
    \If {$op = 0$} 
        \If {$hit = 1$}
            \State{Done}
        \EndIf
        \State {Read $root2$ from SB or root register}
        \State {Forward $root2$ to PA}
    \ElsIf {$op = 1$}
            \State {Read $root2$ from SB or root register}
            \If {$SB\_hit$ = 1}
                \State {Forward $update$ to SB}
                \State {Return $rsp = 1$ to upstream logic}
            \Else
                \State {Forward $update$ to root register}
                \State {Return $rsp = 1$ to upstream logic}
            \EndIf
    \EndIf
\EndIf
\EndWhile
\EndFunction
\end{algorithmic}
\end{algorithm}

Rather, it will overwrite the old version of that node. If the node is not hit in SB, the update logic will search for the requested node in the corresponding BMT cache and eventually memory if required and update the node where it currently resides. As described in the previous section, this technique doesn't fetch any nodes to SB or even BMT cache during update/write. In other words, if $SB\_hit$ and $Cache\_hit$ are both zero, then the controller would directly update the node in the memory instead of bringing it to the cache or SB (algorithm \ref{alg:cap2}). As a result, the MT logic also doesn't have to go through parallel verification (PV) block during node updates. This process is different from HERMES\cite{zou2021hermes} where every request must have a corresponding entry in SB and also must go through PV block during access of unverified nodes, regardless of the counter request type (read/write). 

Another marked difference with the HERMES BMT architecture is the presence of a Write-Back (WB) engine in our design. It allows the write-back of a node during a dirty eviction from a particular level while allowing for other compute levels to work simultaneously. The addition of WB engine makes any relaxed algorithm including HMT to work with this dataflow architecture. Since node verification is done asynchronously, the caching of a buffered node after authentication is decoupled from the ongoing counter operation. As a result, a dirty eviction from the BMT cache due to the caching of a new node might overlap with a current operation on the evicted node. In order to maintain metadata persistence, any write-back request will always get prioritised over an ongoing MT operation. If the eviction takes place in a BMT cache pertaining to the current level, then the node is hashed and its information is relayed to the next MT stage (algorithm \ref{alg:cap2}). Otherwise, if the write-back request comes from a preceding level, then the parent node in the current level is checked and updated using the aforementioned HMT write technique. If the parent is not hit in SB/cache, then the node is hashed and the hash value is conveyed to the next stage until a cached node is arrived upon. As expected from a relaxed update logic, a cache hit in any of the BMT levels would result in skipping all the upper stages and terminate the process in the current stage, no matter whether the counter request corresponds to verification/update. 
\subsection {Root Stage}
The root stage consists of a root SB and the root register access logic. The authentication process involves bringing the root to the SB and then use it for verification of the child node. The update process on the other hand modifies the root on SB or directly in the root register (if not found in the SB). 
\section {Potential Implementation} 
As mentioned in multiple literature (\cite{awad2019triad, awad2019persistently, alwadi2020phoenix, zou2022ares, zubair2019anubis, freij2020persist, chen2020cachetree}), integrity verification is crucial in secure memory-based systems where the contents of the memory is encrypted and the encryption metadata is protected by an integrity tree. 
  \begin{figure*}[h!]
 	\graphicspath{ {./Figures/} }
 	\centering
 	\includegraphics[ scale= 0.4,trim = 0cm 0cm 0cm 0cm]{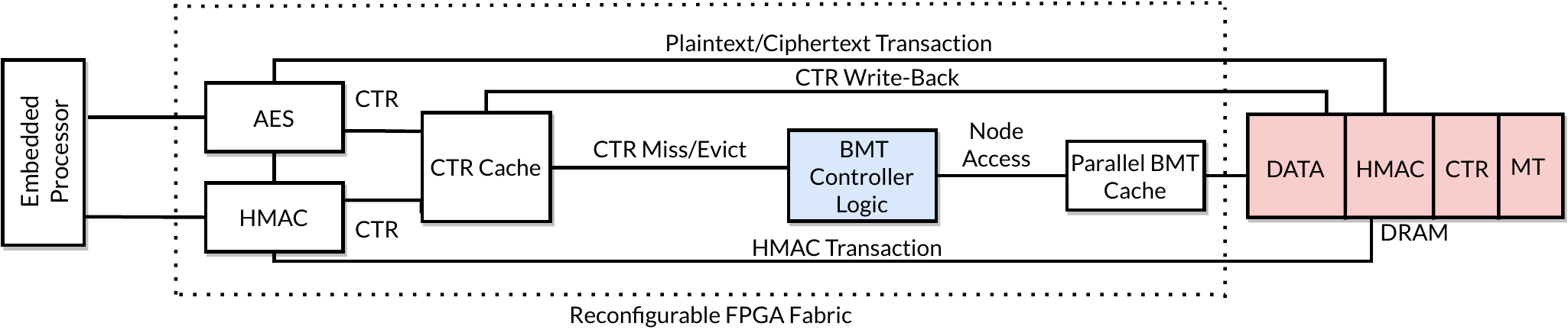}
 	\caption{A FPGA-based secure-memory system where HMT can be utilized}
 	\label{lazyt}
 \end{figure*}
ARES (cite\cite{zou2022ares}) demonstrates the use of such secure memory-based setup in FPGA-driven embedded platform especially with the presence of non-volatile memory as the presence of a hardware-optimized, eager-update BMT is required for atomic recovery of metadata after a system crash. 

Even though a relaxed BMT doesn't support crash-recovery, it can theoretically provide a potential uplift in the run-time system throughput over the eager BMT in write-heavy scenarios. Therefore, a secure FPGA-based embedded system with regular DRAM can benefit from the run-time improvements offered by HMT (Fig.\ref{lazyt}). In addition to the use of AES encryption and HMAC modules, this system also uses an intermediate counter cache to store the verified counters and skips BMT authentication if the counter is hit in the counter cache. Our system, while closely resembling ARES\cite{zou2022ares}, diverges in the sense that it utilizes a volatile DRAM instead of NVM and employs relaxed-update BMT mechanism. 

HMT can facilitate the use of a write-back counter cache since it doesn't need the counter updates to be propagated to the BMT all the time. A write-back counter cache doesn't send any write propagation requests to the BMT during write hit. In this case, the counter cache sends an update request to the BMT only during a dirty counter eviction and it can be overlapped with the regular counter read/write operation. In other words, BMT in this system is only used if there is a read miss in the counter cache or a dirty counter eviction. Hence, HMT can provide additional performance benefits for counter writes. On the contrary, eager-update BMT like the ones used in ARES\cite{zou2022ares} and HERMES\cite{zou2021hermes} must be paired with a write-through counter cache since every counter update must be persisted in the BMT. As the encryption subsystem always requires the previous counter to be fetched first and then incremented and written back during encryption process, a data write operation in all of these systems always comprises of a counter read miss/hit + counter write hit and there is no counter write miss logic used in the counter cache.

\section{Experimental Setup \& Results}
The experimental setup for testing our design includes both subsystem-level and integrated test cases. To test the BMT subsystem isolatedly and measure its peak throughput (Fig.~\ref{lazy9}), we use RSTBenchmark module\cite{zou2021hermes} that is based on the
Repetitive Sequential Traversal (RST) access pattern, commonly used in other recent benchmark suites such as Shuhai \cite{wang2020benchmarking}. The benchmark IP is designed in Verilog RTL and a MicroBlaze soft-core processor is used to control it. The IP generates RST access traces for different stride lengths to emulate the counter requests from a real encryption counter cache. The BMT subsystem consists of BMT controller and our proposed parallel BMT caches. We test the HMT subsystem against a RTL-based lazy update BMT subsystem under the same setup. In other words, both the lazy BMT and HMT controller uses the same cache subsystem. The BMT size for both the BMT subsystem is set to 3 levels and they can protect up to 2MBs of data. Since each tree level has its own cache, they are set to different (1KB L1 + 448BL2 + 64B L3) sizes to have a combined BMT cache size of 1.536KB and all the BMT caches are 4-way set associative write-back caches. The BMT structure under test is a 8-ary tree with 64 bytes of node size. We also test an insecure system that relies on a direct access to the memory without any BMT authentication.

The integrated testing utilizes an embedded secure-memory setup including a memory encryption IP to encrypt/decrypt the application data, a hash-based message authentication code (HMAC) generator for data authentication, a counter cache to store the encryption counters and a BMT subsystem for metadata/encryption counter authentication (fig. \ref{lazyt}). A MicroBlaze processor with a 32 KB L1 cache is used as application driver for benchmarking purposes. In this case, a HMT-based setup is compared against a state-of-the art baseline eager update BMT implementation i.e. HERMES. We use HMT in two different configurations in this system - one with a write-through counter cache mechanism (relaying every write to a cached counter to the BMT) and the other one with a write-back counter caching scheme (only eviction of a dirty counter from the counter cache will deploy a update request to the BMT). On the other hand, the HEMES-based system exclusively uses a write-through counter cache as expected. The BMT size is set to 5-level (metadata corresponding to 128 MB of actual data being protected), the counter cache size is 32 KB and the total BMT cache size used is 40.3 KB (32KB+4KB+4KB+128B+128B) respectively, identical to the ARES system\cite{zou2022ares}. Both the counter cache and BMT caches are direct-mapped caches. The encryption, HMAC and counter cache IPs are sequential in nature and can only process a single request at a time. Aside from the counter cache mechanism and the BMT subsystem, all the other components across all three secure-memory systems remain identical. We use SHA1 hash for the BMT authentication/update in every system.

A Xilinx U200 FPGA accelerator card is used to test all of our designs. 

\subsection {Resource Consumption}

 \begin{table}[h!]
 \begin{threeparttable}
 \begin{tabular}{l c c c c } 
\toprule
  & LUT & Flip-flop & BRAM Tile \\ [0.5ex] 
 \midrule \midrule
  Lazy Update Controller (3L) & 60860(2.8\%) & 65922(3.5\%) & 37.5(1.7\%) \\[0.5ex]
  \hline
  HMT Controller (3L) & 107318(9.1\%) & 115581(4.9\%) & 0(0\%) \\[0.5ex]
  \hline
  HERMES Controller (5L) & 275300(23.3\%) & 316600 (13.4\%) & 29(1.4\%) \\[0.5ex]
  \hline
  HMT Controller (5L) & 328024(27.7\%) & 317274(13.4\%) & 29(1.4\%) \\[0.5ex]
  \hline
  RSTBenchmark IP & 1602(0.1\%) & 1342(0.1\%) & 0(0\%) \\[0.5ex]
 \hline
 BMT Cache (3L) & 8974(0.76\%) & 5711(0.24\%) & 0(0\%) \\ [1ex]
 \hline
 BMT Cache (5L) & 48912(4.1\%) & 40447(1.71\%) & 128(5.9\%) \\ [1ex]
 \bottomrule
\end{tabular}
\caption{Hardware resource utilization numbers for the experimental setup}
\label{table:1}
\end{threeparttable}
\end{table}

The HMT controller incurs more than 40\% higher LUT and similar Flip-flop consumption compared a 3-level lazy BMT. However, this trade-off is justified since the the HMT controller contains notably more complex logic and is tuned for considerably better throughput. Evidently, the lazy BMT uses some BRAM resources while the HMT controller incurs none. The lazy BMT needs the extra BRAM to collect the evicted nodes during a counter verification/update and then write-back at the end of the operation since unlike in HMT, the dirty evictions related to a counter authentication in regular lazy update always happen during that operation. In comparison with the HERMES controller, a 5-level HMT design accounts for additional 19\% LUT and almost similar flip-flop usage. We assume the LUT overhead to be caused by the opportunistic update logic used in HMT as opposed to a uniform eager-update mechanism employed by HERMES.
\subsection{Subsystem Testing}
\begin{figure*}[htbp]
 	\graphicspath{ {./Figures/} }
 	\centering
 	    \begin{subfigure}
 	     \centering
         \includegraphics[scale =0.5,trim = 2cm 0cm 0cm 0cm]{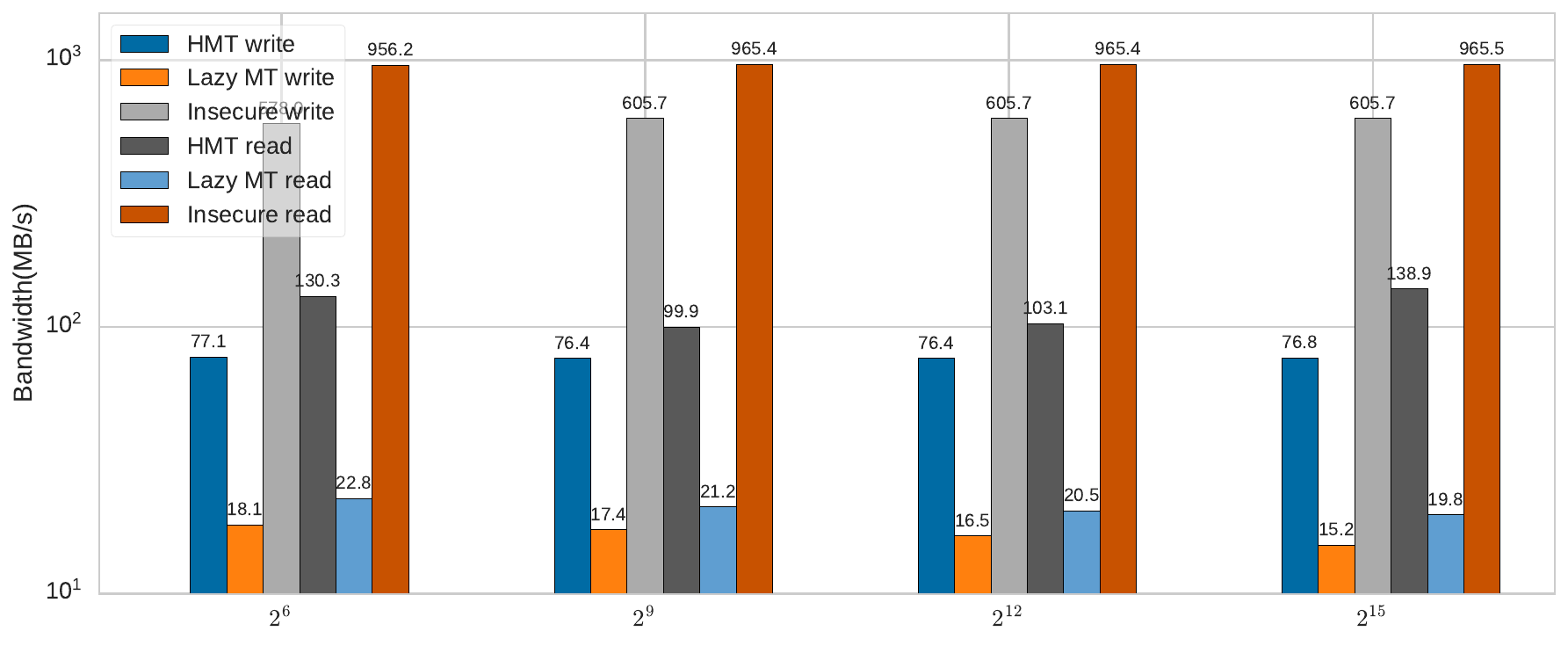}
 	       \label{a}
 	       \end{subfigure}
 	      \begin{subfigure}
 	      \centering
        \includegraphics[scale =0.5,trim = 2cm 0cm 0cm 0cm]{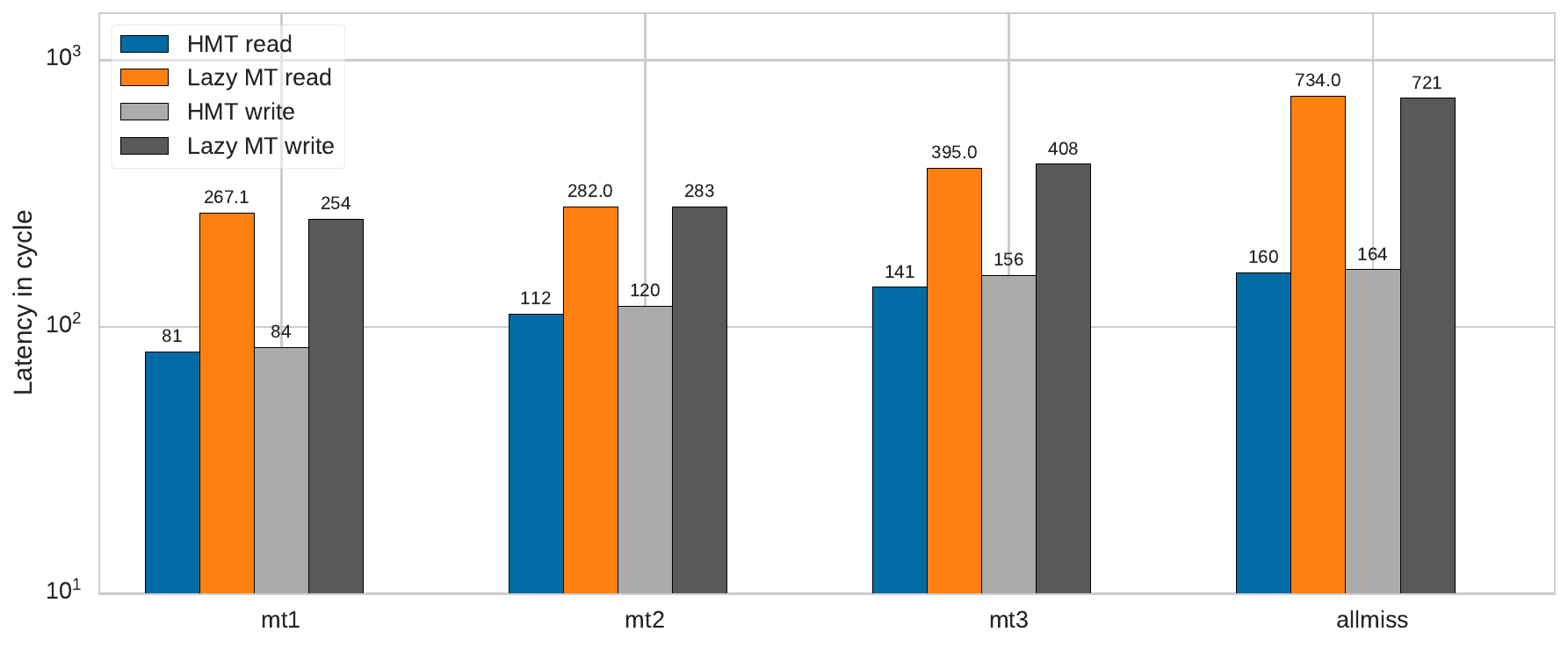}
 	       \label{a}
 	       \end{subfigure}
 	\caption{Read \& write throughput of HMT against the insecure and lazy update system(top), write \& read latency comparison between Lazy update and HMT controller for hits in different levels of BMT caches(bottom)}
 	\label{result3}
 \end{figure*}

As mentioned before, we use variable memory stride lengths in RSTBenchmark to emulate encryption counter access and calculate peak bandwidth of BMT subsystem especially in more randomized access cases. The metadata
locality reduces as the accesses with larger strides ensures more misses in the BMT
caches of lower levels. The HMT controller is essentially a fully pipelined
design and as such, can handle multiple incoming requests even before the
completion of previous authentications. Also, the use of SB ensures speculative
operations based on the nodes currently under authentication which is
especially helpful for applications with good data locality. The lazy BMT, on the other hand, is constrained by its sole operation at a given time. On top of that, it doesn't have the performance enhancements like the use of SB or the reduction in potential evictions due to algorithmic modifications. As a result, HMT demonstrates upto 7x read and 5x more write throughput improvements (Fig. \ref{result3}) over baseline (lazy update).
However, even with these performance enhancements, there is still room for
improvement compared to the bandwidth of the insecure system in Fig.
\ref{result3}. Due to its unique nature of update, we assume that the write
performance of HMT largely relies on BMT cache size instead of stride length of the
access and the constant write bandwidth supports our assumption. On the other hand, the read throughput drops with the higher stride lengths as the HMT encounters more read misses with less spatial locality. However, the stride size of
$2^{15}$ unexpectedly provides better performance and we assume it's due to
additional hits in SB since the system keeps accessing the same two data locations repeatedly.

For latency measurement, we create custom access patterns with specific
addresses that would ensure BMT cache hits in different levels. 'mt0', 'mt1'
and 'mt2' indicates a cache hit in first, second and third level respectively
while 'all Miss' represents a cache miss in every level. We name the BMT
structure in a way such that the first level indicates the parent level of
counter level, the second level is the grandparent level and so on. The
combination of our new HMT algorithm and dataflow architecture results in
noticeable improvement of the latency for both read and write . As each compute
stage in the HMT controller can be activated individually and there's no additional
wait time involved, they can hide the memory access and hash latency
effectively. The read or verification pipeline in HMT controller reduces the
overhead of multiple hash latency to approximately one single hash latency by
leveraging parallel verification module. On the other hand, while the lazy
update controller breaks down the algorithm into different operations, some of
the elements in the datapath remains stalled during particular parts of the
algorithm. Our RTL lazy MT speculatively pre-fetches requests for hiding extra
latency in case of cache misses in multiple levels, but doesn't store the
additional upper level nodes if a lower level hits in the cache. However,
since the immediate child of the root is only 64 bytes in size and has its own
cache, it will never be evicted after it is brought to the cache and therefore,
our lazy BMT does not pre-fetch the root. As a result, in case of 'all miss',
the verification of lower level nodes need to wait for the root authentication
first. The HMT controller, however, handles this scenario in the same way as
any other cache miss and lowers the overhead by approximately 4.5x for both
read and write.
 \begin{figure}[h!]
 	\graphicspath{ {./Figures/} }
 	\centering
 	     \centering
        \includegraphics[scale =0.5,trim = 0cm 0cm 0cm 0cm]{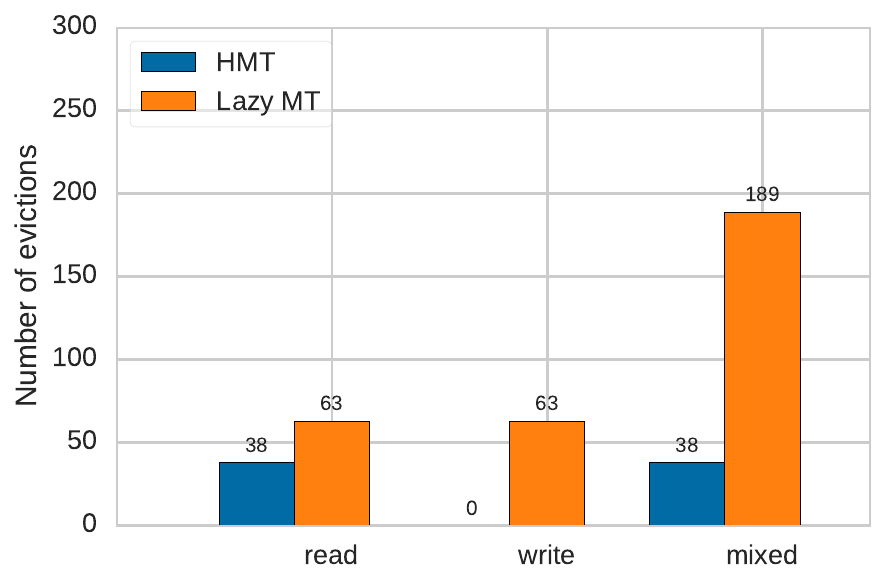}
 	\caption{Number of evictions from BMT cache for different scenario)}
 	\label{resulte}
 \end{figure}
 The performance and latency difference between lazy BMT and HMT subsystems
 depict a notable improvement for the MT authentication due to algorithmic and
 architectural changes in HMT controller. 

In order to compare write-back scenario, We set the stride length of RSTBenchmark to a fixed value of $2^6$ to access every counter and performed counter write followed by a read operation for both lazy BMT and HMT subsystems with a cache size of 2.2KB (2KB L1+ 128B L2 + 128B L3). This setup allows for evictions in a easily repeatable sequential manner as it will give rise to additional secondary evictions caused by the primary evictions in lazy MT. For this test, we consider all evictions from a BMT cache to be dirty evictions. As expected, the counter write results in 0 evictions for HMT compared to 63 for lazy update (fig. \ref{resulte}). This is due to the fact that HMT doesn't require fetching a node to cache or SB on cache miss during a write operation. It also doesn't mandate any integrity check on cache miss during update and directly updates the nodes in the memory until a cached node is found. For counter read, we can see a 40\% reduction in evictions for HMT over the lazy MT. This is also self-explanatory as due to the absence of parent node caching during write-back, the evicted blocks cannot contribute to additional evictions in HMT system even during an eviction due to read operation. As such, the HMT only encounters primary evictions caused by counter verification chain and the worst-case evictions for a HMT operation is only N where N is the number of BMT levels. In the mixed-mode operation where the write is followed by read, the HMT system achieves 5x reduction in number of write-backs due to the gains from both read and write scenarios.

 \begin{figure*}[h!]
 	\graphicspath{ {./Figures/} }
 	\centering
 	     \centering
        \includegraphics[scale =0.5,trim = 0cm 0cm 0cm 0cm]{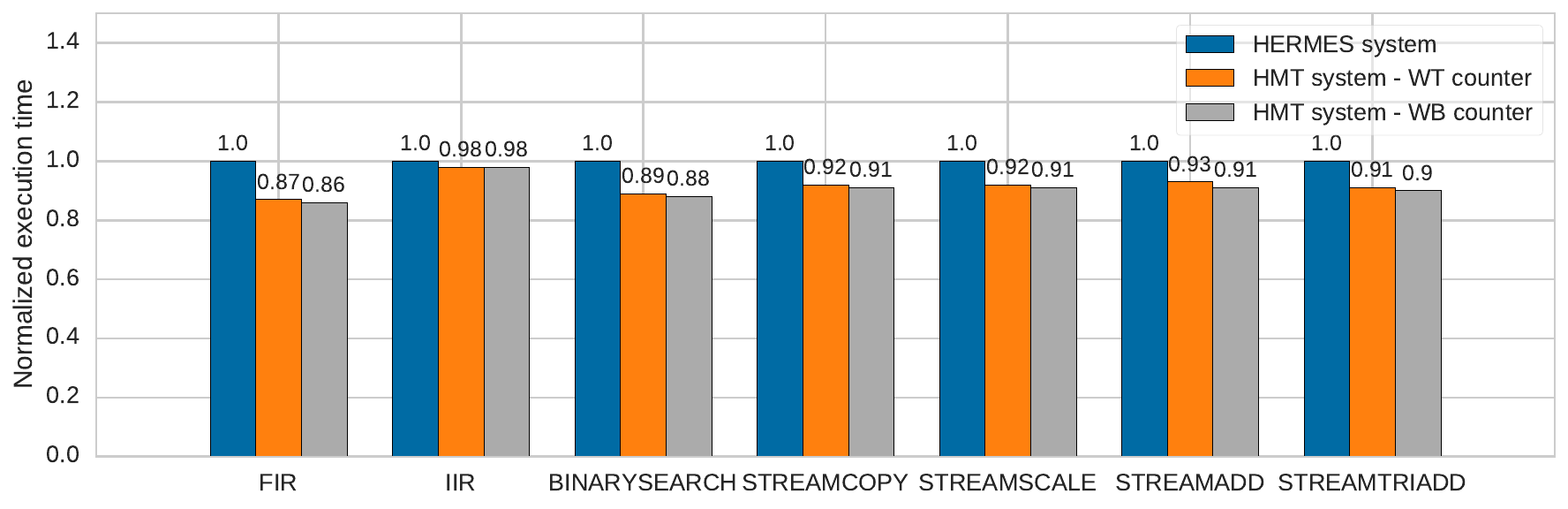}
 	\caption{Performance comparison between HERMES-based and HMT-based systems in standard benchmarks}
 	\label{result4}
 \end{figure*}

\subsection{Integrated Testing}

The overall real-time system read latency for our integrated test setup (fig. \ref{lazyt}) can be characterized by a model, $T\_system = T\_processor + max(T\_encryption, T\_hmac) + max(T\_ctr\_cache, T\_bmt)$ where $T\_processor$ is the latency overhead of processor instructions, $T\_encryption$ is the encryption/decryption latency, $T\_hmac$ is the HMAC generation/verification overhead, $T\_ctr\_cache$ is the counter cache access latency and $T\_bmt$ is the overhead incurred by the BMT subsystem. $T\_bmt$ and $T\_processor$ depends on the workload and the rest of the parameters remain more or less constant throughout the test. Since the HMAC is generated on ciphertext, encryption and HMAC modules can operate in parallel once an encrypted data block is fetched during read operation. Since the test systems are only differentiated by the BMT implementations, the actual performance uplift depends on the number of BMT accesses. For example, if an access request results in a read hit in the counter cache, the BMT will not be employed at all and $max(T\_ctr\_cache, T\_bmt)$ turns into just $T\_ctr\_cache$ which guarantees that all the systems will produce similar results for that particular access. Conversely, a read miss or updates on a cached counter will require BMT authentication/updates and the the total latency will be influenced by $T\_bmt$ instead. As a result, the BMT might be a system bottleneck only when a large number of counter accesses are missed in the counter cache. For write scenario, the latency can be defined by $T\_system = T\_processor + T\_encryption + T\_hmac + max(T\_ctr\_cache, T\_bmt)$ as the HMAC module has to wait for the encryption IP to decrypt the fetched data block and then re-encrypt the modified plaintext into ciphertext before it can generate the new HMAC value. In this case, the system latency is even more dependent on $T\_bmt$ since the BMT overhead can be drastically different across the configurations. For example, HERMES-based setup will propagate every write to the BMT even if the counter is cached and then the update is always persisted all the way up to the root of the tree. HMT with a write-through counter cache will still generate a BMT update request for every write, but the HMT mechanism will stop the update if any tree node in the chain is cached and will not involve any node fetch, resulting in a lower overhead. 

Finally, HMT with write-back counter cache should reduce $T\_bmt$ further by only requiring a tree update during a dirty counter eviction from the counter cache. Since all the IPs other than BMT subsystem are of blocking nature and cannot generate more than one on-flight request at any given time, the BMT cache evictions for HMT will always be performed in the background and won't have any impact on the system latency. It is important to mention that every data write in these integrated systems will result in a counter read (the updated counter is written back to the counter cache and the update is either immediately propagated to the BMT(WT counter cache) or during a dirty counter eviction (WB counter cache)). However, every data read will only result in a counter read.

For integrated system benchmarking, we use STREAM benchmark \cite{mccalpin1995memory,mcvoy1996lmbench}, FIR, IIR benchmarks from DSPStone\cite{zivojnovic1994dspstone} benchmark suite and BINARYSEARCH benchmark from BEEBS\cite{pallister2013beebs} test suite to compare the system level task execution time between baseline and OMT systems. The problem size for all these benchmark are set the same as the configuration in ARES\cite{zou2022ares} (16, 128, 12 \& 128 MB respectively). Every encryption counter protects 4KBs of consecutive data blocks (ciphertext) and therefore, the performance difference among these platforms should increase notably with highly randomized data accesses i.e. higher counter access randomization (as a result, higher BMT utilization) and early write terminations in BMT. The mathematical model developed in HERMES\cite{zou2021hermes} work with the parameters from \cite{zubair2019anubis, ye2018osiris} to approximate the BMT performance in a real-time FPGA-based secure system also supports this notion as it predicts very low BMT overhead for applications with high locality and larger difference in performance for more randomized workloads. Hence we see interesting shifts across benchmarks in fig. \ref{result4} since some of them might exhibit more data locality then others. The IIR workload being highly sequential and compute-intensive, shows very little difference across all three systems. On the other hand, STREAM shows notably better performance in the HMT systems despite being a sequential workload. This difference is due to the fact that STREAM is a write-heavy workload with minor computation overhead and since each write in HERMES system is always propagated to the root of the BMT, the potential early write terminations in HMT allows for lowering of $T\_bmt$. HMT also lowers the execution time by up to 14\% and 12\% respectively in FIR and BINARYSEARCH benchmarks over HERMES. While FIR is more compute-heavy compared to STREAM, it schedules comparatively more randomized memory visits and hence incurs a lot of extra BMT usage which should explain the performance uplift for HMT

Even though BINARYSEARCH exhibits high data locality (according to the analyses in ARES\cite{zou2022ares}), the write access pattern during the key initialization at the start of the benchmark will have notable less overhead for the relaxed update as most of the updates in HMT will stop at the lowest level of BMT cache due to the immediate counter read before the write. HERMES, on the other hand, will have to keep updating every chain to the root and in this case, will find most of the update chains to have at least one BMT cache miss (since the previous read will stop at a cached node and won't cache higher level nodes anymore in that case) and will need to authenticate these nodes first. As a consequence, $T\_bmt$ overhead will be notably reduced.

These benchmark results are in line with the performance model described in HERMES\cite{zou2021hermes} and the latency model discussed above. In all of these cases, the write-back counter caching doesn't provide any real benefits over the write-through counter cache configuration for HMT since a lot of HMT chains should be terminated pretty early due to the cached BMT nodes from earlier operations anyway.

 \begin{figure}[h!]
 	\graphicspath{ {./Figures/} }
 	\centering
 	     \centering
        \includegraphics[scale =0.5,trim = 0cm 0cm 0cm 0cm]{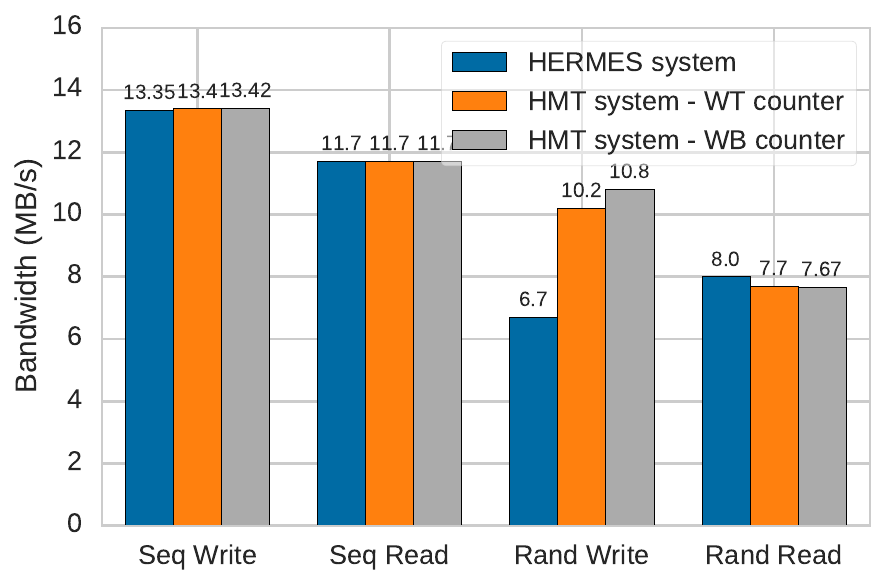}
 	\caption{Bandwidth comparison for read and write accesses}
 	\label{result5}
 \end{figure}

\begin{figure}[h!]
 	\graphicspath{ {./Figures/} }
 	\centering
 	     \centering
        \includegraphics[scale =0.5,trim = 0cm 0cm 0cm 0cm]{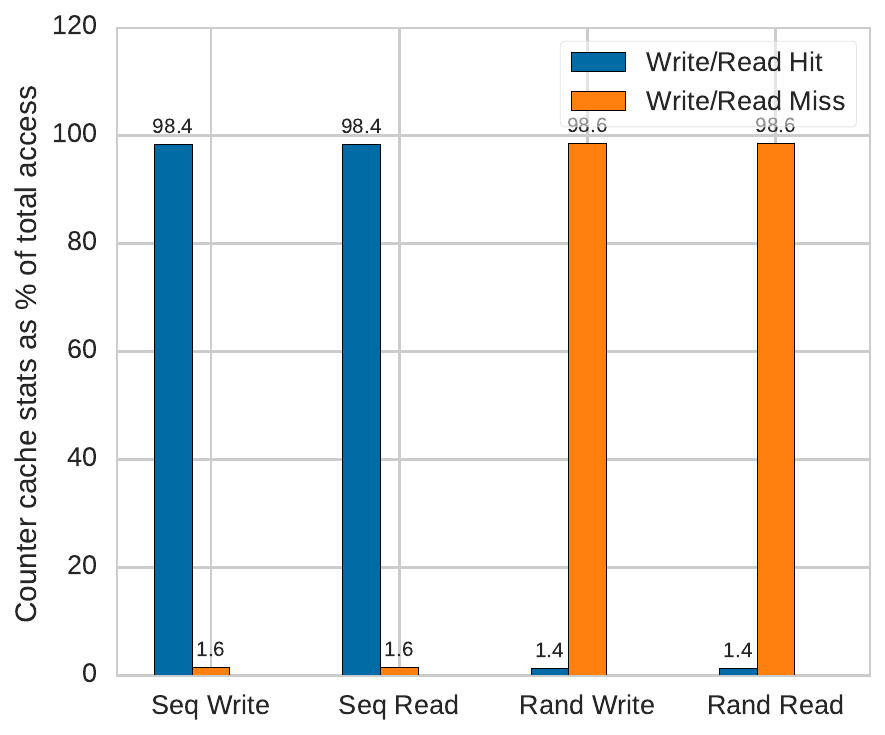}
 	\caption{Counter cache hit and miss frequency}
 	\label{result6}
 \end{figure}

\begin{figure}[h!]
 	\graphicspath{ {./Figures/} }
 	\centering
 	     \centering
        \includegraphics[scale =0.5,trim = 0cm 0cm 0cm 0cm]{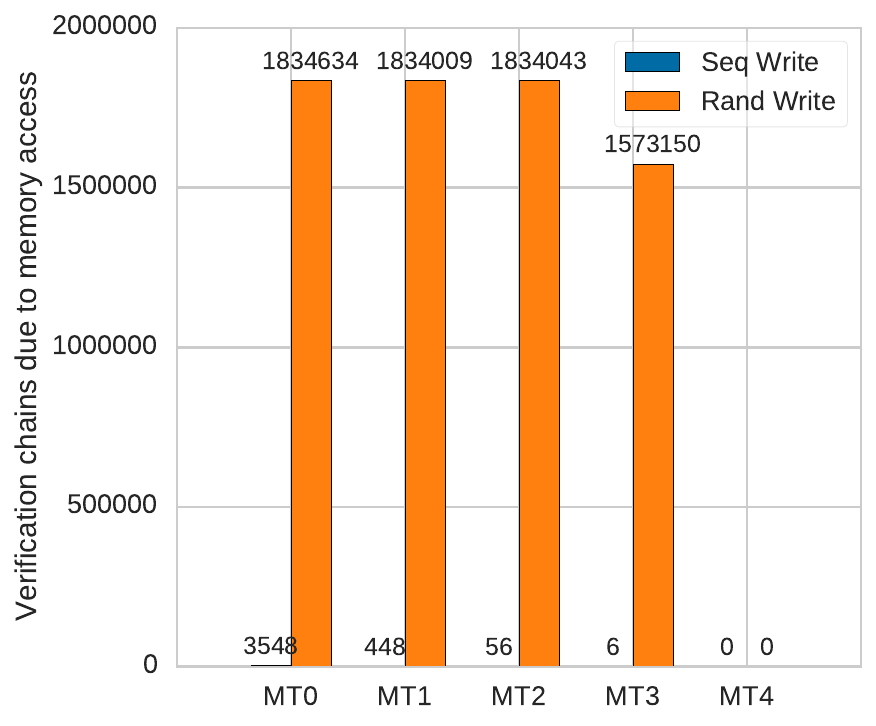}
 	\caption{Authentication overhead for HERMES during write accesses}
 	\label{result7}
 \end{figure}

 As the standard applications used above don't employ BMT in an excessive manner, We traverse through the entire data range with sequential and randomized read \& write operations to saturate the counter access datapath\ref{result5}. Unlike in ARES\cite{zou2022ares}, each of these tests in our platforms are done consecutively (instead of running one at a time) and the access patterns of writes directly affect the following read operations. As seen from fig. \ref{result6}, the sequential read and write accesses incur a hit in the counter cache more than 98\% of time. 
 A write hit should activate the BMT to update the corresponding tree chain for HERMES \& HMT-WT Counter systems (HMT-WB Counter system activates BMT write only during a dirty counter eviction from the counter cache). However, since very few percentage of the preceding counter read are missed for before a counter write (Fig.\ref{result6}), both the relaxed and the eager update BMT in this case visit memory very infrequently and avoid additional verification since most of the update chains will have BMT cache hits in every tree level. As a result, the overall write overhead for eager BMT i.e. HERMES will drop due to low number of verification chains (Fig.\ref{result7}). This unique situation diminishes the performance advantage of relaxed write over eager update as HERMES will return the write response back to the counter cache as soon as it encounters BMT cache hit in all levels for an update chain and not wait for the updates to all the levels to finish, masking most the extra hashing overhead for updates in every level. As a result, even though HMT will terminate the update chain as soon as it reaches the first cached node, the effective overhead reduction will be very minimal compared to HERMES in this case and both the systems produce similar results. We have found HMT in write-back counter configuration in this test to provide very close results to the write-through configuration as the HMT-WT system will have negligible overhead due to high number of BMT cache hits in the first level. As mentioned above, a counter read hit does not require BMT utilization and therefore very few of the counter accesses will go through BMT verification for sequential read, making it also showcase a constant throughput regardless of the BMT implementation. As a result, $T\_bmt$ remains constant across all the systems for the sequential read. 
 The BMT utilization sees a dramatic increase in case of randomized write/read scenario where almost every access reaches BMT for verification/update due to very high percentage of counter miss. In this case, HMT will still experience earlier update terminations due to its update policy as it will still find a cached node before the root/last level in most of the update scenarios. However, HMT overhead will now be noticeable higher compared to the sequential write as most of the updates might no longer have a node hit at the first level. As such, the write-back configuration now further reduces $T\_bmt$ since considerably lower number of writes will need to be propagated to the BMT in HMT-WB system and it results in 6\% uplift in performance. On the other hand, the eager update BMT will need to go through a lot of additional verification chains (Fig.\ref{result7}) since it will need to visit DRAM a lot more frequently and then verify the fetched node first before caching it. As a result, HERMES system will not be able to hide the extra hashing overhead of the updates up to root anymore since this time the BMT controller will need to wait until the end of these verification chains to issue a write response to the counter cache. The extra verification loops and a typically long update overhead increases the $T\_bmt$ and causes the write bandwidth of the baseline to drop by up to 38\% compared to the HMT implementations. The random read, on the other hand, demonstrates a slightly lower read bandwidth for both HMT systems since they suffers from numerous dirty write-back chains due to high number of evictions compared to sequential read. The eager update doesn't need to maintain any write-back update chain and only writes back the evicted block to memory.

\section{Conclusion}

With both algorithmic improvements and hardware innovations, 
we have developed a novel BMT update algorithm to facilitate a
dataflow architecture for BMT verification with intermittent propagation.
This new BMT framework not only improves the performance over the existing
BMT techniques but also makes the design and nature of the BMT propagation very
hardware-efficient and suitable for FPGA-based heterogeneous and embedded
systems. 
Experimental results from both susbsytem and system level tests also provide evidence to our claims and results in up to 14\% better performance in real-time standard benchmarks. We plan to extend the framework to support integration with other state-of-the art BMT
designs in the future.

\appendices
\section{Worst-Case Eviction For Parallel BMT Cache}
Let us consider that the function $E_m$ evicts a dirty block from $m^{th}$ level of BMT, the function $W_m$ writes to $m^{th}$ level \& the function $V_m$ verifies the integrity of the node during memory access and updates the node if required. We assume the worst-case scenario for the eviction, that is - fetching a BMT node from the memory and caching it would cause eviction of a dirty node from the same level as the fetched node (except from the immediate lower level of root) for our proposed cache architecture. Reading or writing to an $m^{th}$ level node requires accessing that node from the memory as the node is not present in the BMT cache. The value $m$ is within $[1,N]$ where $N$ is the number of levels in the BMT (excluding counter and root level) and $1$ denotes the immediate parent level of the counters. Every node accessed from the memory must be authenticated first. Once the integrity is verified, the node might or might not be updated depending on the type of operation (read/write) and put into the cache. As mentioned above, writing the node to the cache would cause a dirty eviction in the worst-case. Therefore, $W_m$ can be written as,
\begin{equation}
W_m \xrightarrow{} E_m + V_m
\label{eq1}
\end{equation} 

However, evicting a dirty cache block from $m^{th}$ level would involve updating the parent node. As a consequence, $E_m$ would be related to $W_{m+1}$,
\begin{equation}
E_m \xrightarrow{} W_{m+1}
\label{eq2}
\end{equation} 

The integrity authentication of an $m^{th}$ level node requires hashing of the corresponding node. The generated hash is compared against the immediate parent node. As part of our worst-case assumption, none of the upper level nodes in the current update/verification chain would be cached beforehand. In other words, the parent would not reside in the cache for the ${m+1}^{th}$ level and caching it would yield another dirty eviction from the same level. In addition, the parent node itself must also be verified and updated as before.
\begin{equation}
V_m \xrightarrow{} E_{m+1} + V_{m+1}
\label{eq3}
\end{equation}
Both eq. \ref{eq1} \& \ref{eq3} can now be modified with the information from eq. \ref{eq2}.
\begin{equation}
W_m \xrightarrow{} W_{m+1} + V_m
\label{eq4}
\end{equation}
\begin{equation}
V_m \xrightarrow{} V_{m+1} + W_{m+2}
\label{eq5}
\end{equation}

Since the highest level of the tree (the level below the root) will have only one node, it will occupy only one cache block. As a result, this node is never going to be evicted under our proposed scheme as there is no other node at the same level to evict it from the cache. In other words, eviction from $N^{th}$ level never takes place, so $E_N$ doesn't exist. Based on this assumption, we can determine the terminal condition from eq. \ref{eq3} when $m = N-1$.
\begin{equation}
V_{N-1} \xrightarrow{} V_{N}
\label{eq6}
\end{equation}
If $m = N$, eq. \ref{eq1} yields,
\begin{equation}
W_N \xrightarrow{} V_N
\label{eq7}
\end{equation}
Using eq. \ref{eq5}, \ref{eq6}, \ref{eq7}  eq. \ref{eq4} can be expanded further. The expanded equation should be our desired expression for the worst-case upper bound.

Therefore, we can write,
\begin{equation*} \label{eq8}
\begin{split}
W_m & \xrightarrow{} W_{m+1} + W_{m+2} + V_{m+1} \\
 & \xrightarrow{} W_{m+1} + W_{m+2} + W_{m+3} + V_{m+2}\\
 & \xrightarrow{} W_{m+1} + W_{m+2} + W_{m+3} + ... V_{N}\\
  & \xrightarrow{} \sum_{k=m+1}^{N} {W_{k}} + V_{N} \\
   & \xrightarrow{} W_{m+1} + \sum_{k=m+2}^{N} {W_{k}} + V_{N}\\
    & \xrightarrow{} \sum_{k=m+2}^{N} {W_{k}} + V_{N} + \sum_{k=m+2}^{n} {W_{k}} + V_{N}\\
    & \xrightarrow{} 2[\sum_{k=m+2}^{N} {W_{k}} + V_{N}]\\
  \end{split}
\end{equation*}
\begin{equation*} \label{eq9}
\begin{split}
 & \xrightarrow{} 2^{2}[\sum_{k=m+3}^{N} {W_{k}} + V_{N}]\\
   & \xrightarrow{} 2^{N-m}[\sum_{k=N}^{N} {W_{k}} + V_{N}]\\
   & \xrightarrow{} 2^{N-m}[W_{N} + V_{N}]\\
   & \xrightarrow{} 2^{N-m+1} V_{N}\\
\end{split}
\end{equation*}

Our cache scheme requires the $N^{th}$ level node to be authenticated only during the first time it is brought from the memory. All the other times, the verification/update chain will stop at $({N-1})^{th}$ level. So we have,
\begin{equation}
W_{m} \xrightarrow{} 2^{(N-1)-m+1} V_{N-1} + V_{N}
\label{eq10}
\end{equation}
 
Here, m = 1 represents update of the immediate parent of a counter during counter write. Therefore, this case includes all the possible recursive chains during a write to counters, including the write-back chains due to evictions. However, the presence of $V^{N}$ in eq. \ref{eq10} represents the original update chain corresponding to the counter write operation. In order to derive the relationship for only the recursions due to dirty evictions, we should remove $V_{N}$ and subtract 1 from the co-efficient $2^{(N-1)-m+1}$ to leave out the original write update chain. Making these changes and using m = 1, we have,
\begin{equation}
W_{m_{Writeback}} \xrightarrow{} (2^{(N-1)} - 1) V_{N-1}
\label{eq11}
\end{equation}

Finally, the total number of worst-case write-back recursive chains or in other words, total number of worst-case evictions due to a counter write for our partitioned cache architecture = $(2^{(N-1)} - 1)$ = $(2^{n} - 1)$ where $n = N-1$. Similarly, it can be proved that a counter read will have the same number of worst-case evictions.

\bibliographystyle{plain} \bibliography{bare_jrnl.bib}
\end{document}